\documentclass[
preprint,
 amsmath,amssymb,
 aps,
]{revtex4-1}

\usepackage{graphicx}
\usepackage{dcolumn}
\usepackage{bm}
\usepackage[colorlinks, linkcolor=blue, anchorcolor=blue, citecolor=blue]{hyperref}
\usepackage{url}
\usepackage{amsmath,amssymb}
\usepackage{newtxmath}
\usepackage[mathlines]{lineno}
\usepackage{subfigure}
\usepackage{eurosym}

\begin{document}


\title{Atom-photon dressed states in a waveguide-QED system with multiple giant atoms}
\author{W. Z. Jia}
\email{wenzjia@swjtu.edu.cn}
\affiliation{School of Physical Science and Technology, Southwest Jiaotong University, Chengdu 610031, China}
\author{M. T. Yu}
\affiliation{School of Physical Science and Technology, Southwest Jiaotong University, Chengdu 610031, China}
\date{\today}

\begin{abstract}
We study the properties of bound states in waveguide-QED systems consisting of multiple giant atoms coupled to a coupled-resonator waveguide. 
Based on the general analytical expressions for these states and the corresponding energy spectra, we analyze in detail the threshold conditions for the appearance of bound states and the photon-mediated interactions between dressed atoms for different configurations. In addition, when multiple giant atoms are coupled to the waveguide, different types of interacting atomic chain can be obtained by manipulating the coupling configurations. Accordingly,
the energy spectra of the bound states form metaband structures in the photonic band gaps. This makes the system a useful platform for quantum simulation and quantum information processing.  
\end{abstract}
\pacs{Valid PACS appear here}
\maketitle


\section{\label{Introduction}Introduction}

Waveguide quantum electrodynamics (wQED) systems \cite{Roy-RevModPhys2017, Gu-PhysicsReports2017}, realized by coupling a single atom or multiple atoms to a one-dimensional (1D) waveguide, 
have attracted widespread attention in recent years. Such systems are excellent platforms for investigating 
strong light-matter interactions at the single-photon level and may have potential applications in 
modern quantum technologies \cite{Shen-OptLett2005, Shen-PRL2005, Chang-PRL2006, Shen-PRL2007, Chang-NaturePhysics2007, Astafiev-Science2010, Longo-PRL2010, Zheng-PRA2010, Fan-PRA2010, Zheng-PRL2011, Bradford-PRL2012, Hoi-PRL2013, Laakso-PRL2014, Shi-PRA2015, Jia-PRA2017, Zhu-PRA2019, Nie-PRAppl2021}. 
The physics of light-matter interactions in 1D becomes even more involved 
when the waveguide is engineered to have finite-bandwidth and nontrivial dispersion relations, e.g., an array of linear or nonlinear optical resonators described by tight-binding model \cite{Zhou-PRL2008, Wang-PRL2020, Scigliuzzo-PRX2022} and a 1D topological photonic bath described by the Su-Schrieffer-Heeger (SSH) model \cite{Bello-ScienceAdvances2019, Kim-PRX2021, Vega-PRA2021}. When quantum emitters are coupled to a finite-bandwidth waveguide, there 
exist atom-photon bound states (BSs) with energies outside the continuum of propagating modes and exponentially localized photonic components \cite{John-PRL1990, John-PRB1991, Tong-JPB2010, Calajo-PRA2016, Shi-PRX2016, Hood-PNAS2016, Liu-NaturePhysics2017, Burillo-PRA2017, Mirhosseini-NatureCommunications2018, Shi-NJP2018, Sundaresan-PRX2019, Roche-PRA2020, Ferreira-PRX2021, Scigliuzzo-PRX2022}, which can be looked on as continuum generalizations of the dressed states in cavity-QED structures \cite{Calajo-PRA2016}. 

With the development of modern nanotechnology, artificial atoms (e.g., transmon qubits \cite{Koch-PRA2007}) can couple to the bosonic modes (phonons or microwave photons) in a 1D waveguide at multiple points spaced wavelength distances, called giant atoms \cite{Kockum-book2021}. 
Such a giant-atom structure can be realized by interacting a transmon qubit with multiple interdigital transducers with surface acoustic waves through piezoelectric effects, or by capacitively coupling an Xmon qubit to a meandering coplanar waveguide \cite{Kockum-PRA2014}.
Unlike point-like small atoms, the usual dipole approximation no longer works for giant atoms, and so the interference and time-delay effects lead to some novel phenomena, such as frequency-dependent decay rate and Lamb shift \cite{Kockum-PRA2014, Kannan-Nature2020, Vadiraj-PRA2021}, decoherence-free interaction \cite{Kockum-PRL2018, Kannan-Nature2020, Du-PRA2023}, non-Markovian dynamics \cite{Guo-PRA2017, Andersson-NaturePhysics2019}, and unconventional scattering spectra due to effects beyond the dipole approximation \cite{Guo-PRA2017, Ask-Arxiv2020, Cai-PRA2021, Feng-PRA2021, Zhu-PRA2022, Yin-PRA2022}. The effects of chiral atom-waveguide coupling \cite{Soro-PRA2022} and ultra strong coupling \cite{Noachtar-PRA2022, Terradas-Brianso-PRA2022} in the wQED systems with giant atoms have also been investigated. 
The structure of giant atom can also be implemented with cold atoms in optical lattices \cite{Tudela-PRL2019}, with a giant spin ensemble \cite{Wang-NatureCommunications2022}, or even in a synthetic dimension \cite{Du-PRL2022, Xiao-NPJ2022}.
The light-matter interactions for giant atoms coupled to a finite bandwidth waveguide have also received attention. An important effect in this kind of structure is atom-photon BS \cite{Zhao-PRA2020, Cheng-PRA2022,Soro-PRA2023,Wang-PRL2021}, which can be manipulated by using interference effects compared with their counterparts in small-atom setups. 

In this paper we analyze the properties of BSs in a system of multiple giant atoms 
coupled to a common photonic bath realized by a 1D array of coupled resonators. Specifically, for the cases of single- and double-giant-atom BSs, the threshold conditions for the BS and the dipole-dipole-like interactions mediated by the BS photons for different configurations are analyzed in detail. Based on these results, we further investigate the multi-giant-atom BSs and find that different type of atomic chain, e.g., the SSH chain, can be constructed by designing the layout of the connection points as well as the interaction strengths at these points. The corresponding energy spectra form metaband structures in the photonic bandgaps, which can be well explained in terms of effective model described by interacting dressed-atom chain. These results may have potential applications in quantum simulation and quantum information processing.
 
The remainder of this paper is organized as follows. In Sec.~\ref{GeneralResultBS} we provide general analytical expressions for the single-excitation atom-photon BSs. In Secs.~\ref{BS-one-GA}, \ref{BS-two-GA}, and \ref{BS-many-GA}, we discuss the properties of single-, double-, and multi-giant-atom BSs, respectively. Finally, further discussions and conclusions are given in Sec.~\ref{conclusion}.
\section{\label{GeneralResultBS}General analytical expressions for the atom-photon bound states}
\begin{figure}[t]
\centering
\includegraphics[width=0.6\textwidth]{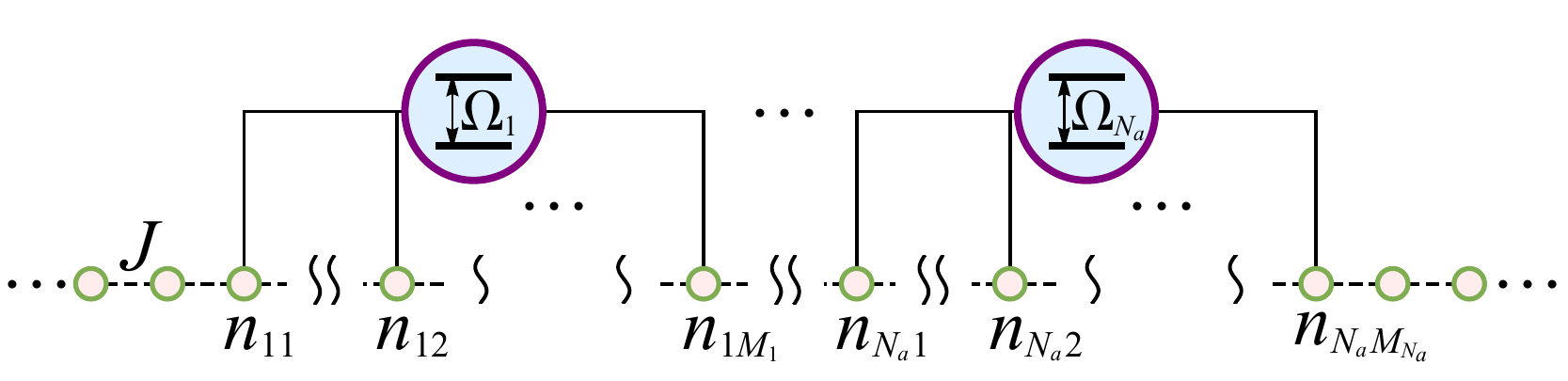}
\caption{A schematic of multiple giant atoms coupled to an array of coupled optical resonators.}
\label{ManyGsys}
\end{figure}

We consider a system where a set of $N_{\mathrm{a}}$ two-level giant atoms is coupled to a finite-bandwidth optical waveguide. The waveguide is modeled as an array of $N\to\infty$ optical resonators with center frequency $\omega_c$ and a nearest-neighbor tunnel coupling $J$. The $m$th ($m=1,2, \cdots, N_{\mathrm{a}}$) atom couples to $M_m$ resonators simultaneously, as shown in Fig.~\ref{ManyGsys}. 
The total Hamiltonian for this system is ($\hbar=1$)
\begin{eqnarray}
H=
\omega_{c}\sum_j c_j^\dagger  c_j - J \sum_j \left( c_j^\dagger  c_{j+1} +\text{H.c.} \right)  + \sum_{m=1} ^{N_{\mathrm{a}}} \Omega_{m}\sigma_{m}^{+}\sigma_{m}^{-} + \sum_{m=1} ^{N_{\mathrm{a}}} \sum_{l=1} ^{M_m} g_{ml} \left( c_{n_{ml}}^\dagger  \sigma_{m}^{-} + \text{H.c.} \right),
\label{SysHamiltonian}
\end{eqnarray}
where $c_j$ ($c_j^\dagger$) is the annihilation (creation) operator of the photonic mode at the $j$th site. $\sigma_m^{+}=|e\rangle_m\langle g|$ ($\sigma_m^{-}=|g\rangle_m\langle e|$) is the raising (lowering) operator of the $m$th atom. $\Omega_{m}$ is the transition frequency between the ground state $|g \rangle_m$ and the excited state $|e \rangle_m$. $n_{ml}$ is used to label the site connecting to the $l$th coupling point of the $m$th atom, the corresponding atom-photon coupling strength is $g_{ml}$. Note that in Eq.~\eqref{SysHamiltonian} we have performed rotating wave approximation by assuming that $\omega_c\simeq\Omega_m$ and that $\omega_c,\Omega_m\gg J, g_{ml}$.

By introducing momentum operators $c_{k}=\frac{1}{\sqrt{N} } \sum_{j} e^{-ikj} c_{j}$,
with $k \in \left[ -\pi, \pi \right]$, we can obtain the Hamiltonian in momentum space
\begin{eqnarray}
H=\sum_k \omega_{k} c_k^\dagger  c_k + \sum_{m=1}^{N_{\mathrm{a}}} \Omega_m \sigma_{m}^{+}\sigma_{m}^{-}
+ \frac{1}{\sqrt{N}} \sum_k \sum_{m=1}^{N_{\mathrm{a}}} \sum_{l=1}^{M_m} g_{ml} \left( e^{-ikn_{ml}} c_{k}^{\dagger} \sigma_{m}^{-}+\text{H.c.} \right),
\label{SysHk}
\end{eqnarray}
where the tight-binding Hamiltonian of the resonator array is rewritten in the diagonal form, with mode frequencies
$\omega_k = \omega_c - 2J \cos k$ lying inside a band with central frequency $\omega_c$ and total width $4J$. 

In the single-excitation subspace, an eigenstate of Hamiltonian \eqref{SysHk} has the form
\begin{equation}
\left|\psi\right> = \left(\cos\theta\sum_{m=1}^{N_{\mathrm{a}}} u_{m}\sigma_{m}^{+}+\sin\theta\sum_{k}f_{k}c_{k}^{\dagger}\right)\left|G\right\rangle,
\label{wavefunction}
\end{equation}
where $u_m$ and  $f_k$ are the normalized excitation amplitudes of the $m$th atom and a photon with wave vector $k$, respectively,  $\theta$ is the mixed angle of the atomic and the photonic components. $\left| G \right>$ is the ground state of system. 
Plugging this ansatz into the eigen equation $H\left|\psi\right> = E \left| \psi \right>$ and eliminating the degrees of freedom of the waveguide modes,
yields the following equations
\begin{equation}
\mathbf{H}\mathbf{u}=E\mathbf{u},
\label{EqForU}
\end{equation}
where $\mathbf{u}=(u_1,u_2,\cdots,u_{N_{\mathrm{a}}})^{\top}$ and
\begin{equation}
\mathbf{H}=\mathrm{diag}(\delta_1,\delta_2,\cdots,\delta_{N_{\mathrm{a}}})+\mathbf{\Sigma}(E).
\end{equation}
Here we have changed into a rotating frame with respect to $\omega_c$ (i.e., making the substitution $E-\omega_{c}\to E$). $\delta_{m}=\Omega _m - \omega_c$ is the atom-cavity detuning. The elements of the energy correction matrix $\mathbf{\Sigma}(E)$ are defined as
\begin{equation}
{\Sigma}_{mm'} \left( E \right)=\sum_{l=1}^{M_m} \sum_{l'=1}^{M_{m'}}\frac{g_{ml} g_{m'l'}}{2\pi} \int_{-\pi}^{\pi} \mathrm{d}k\frac{e^{ik\left(n_{ml}-n_{m'l'}\right)}}{E+2J\cos k}.
\label{ELcorrection}
\end{equation}
The diagonal element $\Sigma_{mm}(E)$ represents the self-energy correction of the atom $m$ due to  emission and re-absorption of photons by the same atom. The off-diagonal term $\Sigma_{mm'}(E)$ ($m\neq m'$) represents the mutual-energy correction due to photon exchange between atoms $m$ and $m'$ through the optical waveguide, which can lead photon-mediate interactions between atoms.
The integral in Eq.~\eqref{ELcorrection} is calculated explicitly \cite{Economou-QuantumPhysics1979} using $|E| > 2J$:   
\begin{equation}
{\Sigma}^{(\beta)}_{mm'} \left( E \right)=\frac{\sum_{l=1}^{M_m} \sum_{l'=1}^{M_{m'}} g_{ml} g_{m'l'} (-\beta)^{\left| n_{ml}-n_{m'l'} \right|} e^{-\frac{ \left| n_{ml}-n_{m'l'} \right|}{\lambda(E)} }}{E \sqrt{1-\frac{4J^2}{E^2}}},
\label{ELcorrectionA}
\end{equation}
with $\lambda(E)$ being defined as
\begin{equation}
\lambda(E)=\left[\mathrm{arccosh}\left(\frac{\left|E\right|}{2J}\right)\right]^{-1}.
\label{LL}
\end{equation}
When $\beta=1$ ($\beta=-1$), the expression \eqref{ELcorrectionA} represents the energy-correction function for $E>2J$ ($E<-2J$), which can be used to determine the energy of the BS above (below) the scattering continuum. 

One can see from Eq.~\eqref{EqForU} that the energies of the BSs are the real solutions of the following equation
\begin{equation}
\det [\mathbf{H}- E\mathbf{I}] = 0,
\label{BSeq}
\end{equation}
with $\mathbf{I}$ being the identity matrix. Note that this equation is a transcendental equation because $\mathbf{H}$ is a function of $E$. For fixed $\beta$, there are at most $N_{\mathrm{a}}$ real solutions, labeled as $E_{\beta s}$ [$s=1,2,\cdots s_{\mathrm{m}}$ ($s_{\mathrm{m}}\le N_{\mathrm{a}}$)]. And when $\beta=1$ ($\beta=-1$), the solutions satisfy $E_{1 s}>2J$ ($E_{-1 s}<-2J$), corresponding to the energies of the upper (lower) BSs.

After fixing the energy $E_{\beta s}$ of a BS, one can further obtain the corresponding excitation amplitudes $u^{(\beta s)}_{m}$ of the atoms from Eq.~\eqref{EqForU}. The corresponding expression of BS in real space can be written as 
\begin{equation}
\left|\psi_{\beta s}\right>=\sum_{m=1}^{N_{\mathrm{a}}}u^{(\beta s)}_{m}D_{\beta s}^{\dagger}(m)\left|G\right>,
\label{DressState}
\end{equation}
where the dressed-state creation operator $D_{\beta s}^{\dagger}(m)$ related to atom $m$ is defined as
\begin{equation}
D^{\dagger}_{\beta s} (m) = \cos \theta_{\beta s} \sigma_{m}^{+}+\beta\sin \theta_{\beta s} \sum_{l=1}^{M_m} \frac{\tilde{c}^{\dagger}_{\beta s}(n_{ml})}{\mathcal{N}_{\beta s}}.
\label{DressOperator}
\end{equation}
The unnormalized photonic creation operator takes the form
\begin{equation}
\tilde{c}^{\dagger}_{\beta s}(n_{ml}) =\sum_{j} \tilde{g}_{ml} (-\beta)^{ \left| j-n_{ml} \right|} e^{-\frac{\left| j-n_{ml} \right|}{\lambda_{\beta s}}} c_{j}^{\dagger},
\end{equation}
which creates a photon in an exponentially localized wave packet around the $l$th coupling point of the $m$th atom, with localization length being defined as $\lambda_{\beta s}=\lambda(E_{\beta s})$. The mixing angle $\theta_{\beta s}$ satisfies the relation
$\tan \theta_{\beta s} = {\mathcal{N}_{\beta s}}/{(2 \sinh \frac{1}{\lambda_{\beta s}})}$.
The normalization constant $\mathcal{N}_{\beta s}$ is given by
\begin{equation}
\mathcal{N}_{\beta s}=\sqrt{\sum_{mm'}\sum_{ll'}u^{(\beta s)}_{m}u_{m'}^{(\beta s)*}\tilde{g}_{ml}\tilde{g}_{m'l'} (- \beta)^{d_{ml,m'l'}} \left( \coth {\lambda_{\beta s}^{-1}} + d_{ml,m'l'} \right) e^{-\frac{d_{ml,m'l'}}{\lambda_{\beta s}}}},
\end{equation}
where $\tilde{g}_{ml}={g_{ml}}/{J}$ is the coupling strength scaled by $J$. $d_{ml,m'l'}=\left| n_{ml} - n_{m'l'} \right|$ is the distance between the coupling points.
From E.q.~\eqref{DressState}, one can see that  the photonic excitation amplitude at site $j$ takes the form $\tilde f^{(\beta s)}_{j}=\beta\sin \theta_{\beta s}f^{(\beta s)}_{j}$, with
\begin{equation}
f^{(\beta s)}_{j} = \frac{1}{\mathcal{N}_{\beta s}}\sum_{m=1}^{N_{\mathrm{a}}}\sum_{l=1}^{M_m}u^{(\beta s)}_m\tilde{g}_{ml}(-\beta)^{ \left|j-n_{ml}\right|}e^{-\frac{\left| j-n_{ml}\right|}{\lambda_{\beta s}}}.
\label{PhotonicAmplitude}
\end{equation}

The BS obtained here are atom-photon dressed state with energy in the band 
gap and photonic component exponentially localized around each coupling point.
Note that in the wQED systems with giant atoms, there also exist another kind of BS in the continuum with photons confined between different coupling points \cite{Guo-PRR2020, Guo-PRA2020}, and similar phenomena have been studied previously in wQED systems with small atoms \cite{Tufarelli-PRA2013, Calajo-PRL2019}. 
\section{\label{BS-one-GA}Bound states for a single giant atom}
Let us first consider a single giant atom with $N_{\mathrm{c}}$ connection points. The $l$th coupling point connects to the site $n_{l}$ with
coupling strength $g_{l}$. In this case, equation \eqref{BSeq} reduces to the following form
\begin{equation}
E-\delta=\Sigma_\beta(E),
\label{BSeqnSG}
\end{equation}
with
\begin{equation}
\Sigma_\beta(E)=\frac{\sum_{l,l'=1}^{N_{\mathrm{c}}} g_l g_{l'} (-\beta)^{\left|n_l-n_{l'}\right|}e^{-\frac{\left|n_l -n_{l'}\right|}{\lambda(E)}}}{E \sqrt{1- \frac{4J^2}{E^2}}}
\label{energyequationSG}
\end{equation}
being the self-energy function of a single giant atom. $\delta$ is the detuning between the transition frequency of the atom and the central frequency of the waveguide. We label the real solutions of Eq.~\eqref{BSeqnSG} as $E_\beta$, with $\beta=\pm 1$ representing the upper and lower BSs, respectively.
The corresponding wave functions can be written as $\left|\psi_{\beta}\right>=D_{\beta}^{\dagger}\left|G\right>$,  where $D_{\beta}^{\dagger}$ is the single-atom dressed-state creation operator.

\begin{figure}[t]
\centering
\includegraphics[width=\textwidth]{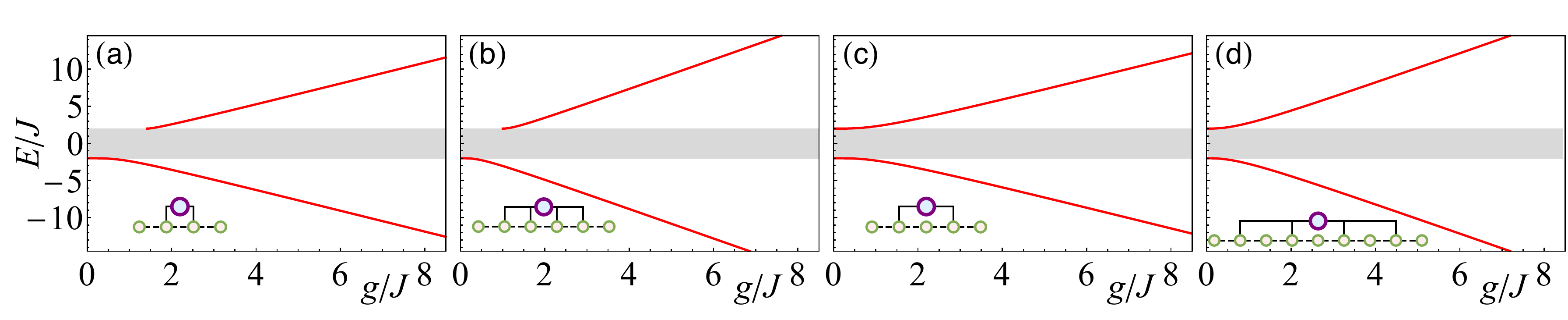}
\caption{The BS energy levels for the case of a single giant atom are plotted as functions of the coupling strength $g$ for different values of $\Delta n$ and $N_{\mathrm{c}}$. The inset in each panel shows the corresponding schematic of system. For all plots $\delta=0$ is assumed.}
\label{Spectrum1Atom}
\end{figure}

Although the BSs for a single giant atom were numerically investigated previously \cite{Zhao-PRA2020}, we can extract more properties of the BSs from 
above analytical expressions. Based on these one can further investigate the BS for double and multiple giant atoms (see Secs.~\ref{BS-two-GA} and \ref{BS-many-GA}). We first consider the threshold conditions of the BS. As shown by Eq.~\eqref{BSeqnSG}, the energy of the BS can be determined by 
the intersection points of the curves $y=E-\delta$ and $y=\Sigma_\beta(E)$. Note the self energy $\Sigma_{\pm 1}(E)$ 
monotonically decreases with $E$ when $|E|>2J$. And for a single small atom, $\Sigma_{\pm 1}(\pm 2J)$ diverges to $\pm\infty$, thus a single BS of 
energy $E_{-1}<-2J$ ($E_{1}>2J$) certainly occurs for all $g$ \cite{Calajo-PRA2016}. However, for a single giant atom, due to the non-dipole effects (i.e., 
photon exchange between different coupling points), the self energy at the bandedge may be finite under some parameters. Thus the appearance of the 
upper BS requires a coupling strength larger than a threshold value. To show this, we focus on the special case that the distance between neighboring 
coupling points is a constant $\Delta n$ and the coupling strengths are all the same, denoted as $g$. Under this configuration, one can find a finite self 
energy $\Sigma_1(2J)=g^2 N_{\mathrm{c}}\Delta n/(2J)$ at the upper bandedge for $N_{\mathrm{c}}\in\mathbb{E}^{+}$ and $\Delta n\in\mathbb{O}^{+}$. 
Thus, when $N_{\mathrm{c}}\in\mathbb{E}^{+}$, $\Delta n\in\mathbb{O}^{+}$ and $\delta<2J$, to ensure that the linear function $y=E-\delta$ and the 
monotonically decreasing function $y=\Sigma_1(E)$ have an intersection point at $E>2J$, the condition $2J-\delta<\Sigma_1(2J)$ is required, from 
which one can read the threshold condition $g>\sqrt{{2J(2J-\delta)}/{(N_{\mathrm{c}}\Delta n)}}$ for the upper BS. While for coupling strength below this 
value, the BS disappears, as shown in Figs.~\ref{Spectrum1Atom}(a) and \ref{Spectrum1Atom}(b). Moreover, (i) when $N_{\mathrm{c}}\in\mathbb{E}^{+}$, $\Delta n\in\mathbb{O}^{+}$ and $\delta>2J$, $2J-\delta$ is negative while $\Sigma_1(2J)$ is positive; (ii) when $N_{\mathrm{c}}$ and $\Delta n$ take other values, we obtain $\Sigma_1(2J)\to+\infty$. Thus, for these two cases, the value taken by the linear function $y=E-\delta$ at $E=2J$ is always smaller than $\Sigma_1(2J)$, which means that an upper BS of energy $E>2J$ certainly occurs for any values of $g$. Similar analysis shows that $\Sigma_1(-2J)\to-\infty$ is always satisfied, thus a BS below the continuum of energy $E<-2J$ always exists (see Fig.~\ref{Spectrum1Atom}). 

In the strong coupling limit $g\gg J$, the photons are almost confined in the resonators connecting to the atom. The system thus can be looked on as a cavity-QED system with an atom simultaneously coupled to $N_{\mathrm{c}}$ non-interaction cavities, or equally, an atom coupled to a cavity with strength $\sqrt{N_{\mathrm{c}}}g$. Accordingly, the BSs can be approximated as $\big(\cos\theta_{\beta}\sigma^{+}+\beta\sin\theta_{\beta}\sum_{l=1}^{N_{\mathrm{c}}}{{c}^{\dagger}_{n_l}}/{\sqrt{N_{\mathrm{c}}}}\big)|G\rangle$ with corresponding energies $E_{\beta}\simeq \delta/2+{\beta}\sqrt{\delta^2+4N_{\mathrm{\mathrm{c}}} g^2}/2$ (e.g., for $\delta=0$, $E_{\beta}\simeq\beta\sqrt{N_{\mathrm{c}}}g$, as shown in Fig.~\ref{Spectrum1Atom}). The mixing angle becomes $\tan2\theta_{\beta}=2\beta\sqrt{N_{\mathrm{c}}}g/\delta$, satisfying $\cos\theta_{\beta}=\sin\theta_{-\beta}$.

\section{\label{BS-two-GA}Bound states for double giant atoms}
Now we consider the case of two atoms (denoted by $a$ and $b$). According to Eq.~\eqref{BSeq}, we obtain the transcendental equation for the energy of BS
\begin{equation}
E = \frac{1}{2} \left(\tilde{\delta}^{(\beta)}_a(E) +\tilde{\delta}^{(\beta)}_b(E) +\zeta\sqrt{[\tilde{\delta}^{(\beta)}_{ab}(E)]^2 + 4 [\Sigma_{ab}^{(\beta)}(E)]^2 } \right),
\label{BSeqnDoubleG}
\end{equation}
with $\zeta=\pm 1$, $\tilde{\delta}^{(\beta)}_{m}(E)=\delta_{m}+\Sigma^{(\beta)}_{mm}(E)$ ($m=a,b$) and
$\tilde{\delta}^{(\beta)}_{ab}(E)=\tilde{\delta}^{(\beta)}_a(E)-\tilde{\delta}^{(\beta)}_b(E)$, respectively. $\Sigma_{ab}^{(\beta)}(E)$ is the mutual energy between the atom $a$ and $b$. 
From  Eq.~\eqref{BSeqnDoubleG}, one can fix at most four real solutions, labeled as $E_{\beta\zeta}$.
The corresponding atomic excitation amplitudes can be obtained from Eq.~\eqref{EqForU} and take the form
\begin{equation}
u^{(\beta\zeta)}_a =\sin\Theta_{\beta\zeta},~~~u^{(\beta\zeta)}_b=\cos\Theta_{\beta\zeta},
\label{uaub}
\end{equation}
with the mixing angle
\begin{equation}
\tan\Theta_{\beta\zeta}=\frac{-2 \Sigma^{(\beta)}_{ab}(E_{\beta\zeta})}{\tilde{\delta}^{(\beta)}_{ab}(E_{\beta\zeta})
- \zeta \sqrt{[\tilde{\delta}^{(\beta)}_{ab}(E_{\beta\zeta})]^2 + 4[\Sigma^{(\beta)}_{ab}(E_{\beta\zeta})]^2}}.
\label{MixAngTheta}
\end{equation}
The corresponding atom-photon BSs in real space can be expressed by Eq.~\eqref{DressState} by letting $m=a,b$ and $s=\zeta$.

\begin{figure*}[t]
\includegraphics[width=\textwidth]{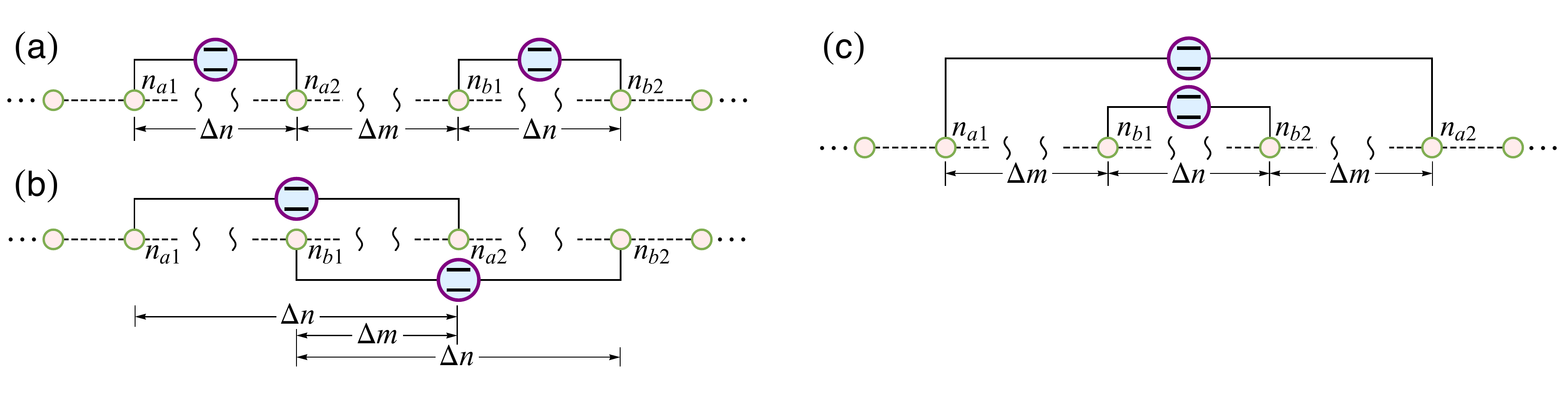}
\caption{Sketches of two giant atoms coupled to an array of coupled optical resonators for three distinct topologies: (a) two separate giant atoms, (b) two braided giant atoms, and (c) two nested giant atoms.
}
\label{TwoGsys}
\end{figure*}

The above results are applicable to any configuration containing double giant atoms. In what follows, we focus on the special case that the frequencies of the two atoms are equal (i.e., $\delta_{a}=\delta_{b}=\delta$), each of the atoms has two connection points, and all the coupling strengths are identical  ($g_{ml}=g$). We also assume that the coupling points are symmetrically distributed (see Fig.~\ref{TwoGsys}), thus each BS has definite parity. The configurations shown in Figs.~\ref{TwoGsys}(a)-\ref{TwoGsys}(c) are called separate, braided, and nested giant atoms \cite{Kockum-PRL2018}, respectively. 
\subsection{\label{EnergyCorrection} Energy correction functions for three configurations}
In this subsection, we provide analytical expressions for the energy correction functions for the configurations shown in Fig.~\ref{TwoGsys}.  
First we consider the configuration containing two separate or braided atoms. We can let $\Delta n=n_{a2} - n_{a1}=n_{b2}-n_{b1}$ and $\Delta m=|n_{b1}-n_{a2}|$, as shown in Figs.~\ref{TwoGsys}(a) and \ref{TwoGsys}(b). The self energies of the two atoms are identical:
\begin{equation}
\Sigma^{(\beta)}_{aa}(E)=\Sigma^{(\beta)}_{bb}(E) =\frac{2g^2 \left(1+ (-\beta)^{\Delta n} e^{-\frac{\Delta n}{\lambda(E)}}\right)}{E\sqrt{1-\frac{4J^2}{E^2}}}\equiv\tilde\Sigma_\beta(E,\Delta n).
\label{SelfEnergySB}
\end{equation}
And the mutual-energy becomes
\begin{equation}
\Sigma^{(\beta)}_{ab} (E) =
\left\{
\begin{array}{ll}
{g^2(-\beta)^{\Delta m} e^{-\frac{\Delta m}{\lambda(E)}} \left( 1+ (-\beta)^{\Delta n} e^{-\frac{\Delta n}{\lambda(E)}}\right)^2}\left({{E\sqrt{1-\frac{4J^2}{E^2}} }}\right)^{-1},~~~~~\mathrm{separated} 
\\
{g^2 (-\beta)^{\Delta m} e^{-\frac{\Delta n-\Delta m}{\lambda(E)}}\left(e^{\frac{\Delta n-2\Delta m}{\lambda(E)}}+ e^{-\frac{\Delta n}{\lambda(E)}}+ 2(-\beta)^{\Delta n}\right)}\left({{E\sqrt{1-\frac{4J^2}{E^2}} }}\right)^{-1},~~~~~\mathrm{braided} 
\end{array}
\right.
\label{MutualEnergyS}
\end{equation}
For present case of two identical atoms with $\tilde{\delta}^{(\beta)}_{ab}(E)=0$, equation \eqref{BSeqnDoubleG} becomes
\begin{equation}
E-\delta=\tilde\Sigma_\beta(E,\Delta n)+\alpha\Sigma^{(\beta)}_{ab}(E). 
\label{energyequationP}
\end{equation}
This equation has up to four solutions outside the scattering continuum, denoted as $E_{\beta\alpha}$. The corresponding atomic excitation amplitudes become $u^{(\beta\alpha)}_a={1}/{\sqrt{2}}$ and $u^{(\beta\alpha)}_b={\alpha}/{\sqrt{2}}$. Here $\alpha=\zeta\mathrm{sgn}\big({\Sigma^{(\beta)}_{ab}}\big)=\pm 1$ is used to label the even- and odd-parity states, respectively. According to Eq.~\eqref{PhotonicAmplitude}, the corresponding photonic wave functions $f^{(\beta,\pm 1)}_j$ are even (odd) functions of $j$.

For the nested configuration shown in Fig.~\ref{TwoGsys}(c), we assume $\Delta n=n_{b2}-n_{b1}$ and  $\Delta m=n_{b1}- n_{a1}=n_{a2}-n_{b2}$. 
Clearly, the two atoms are not identical for this configuration, and the self-energies are $\Sigma^{(\beta)}_{aa} (E)=\tilde\Sigma_{(\beta)}(E,\Delta n+2\Delta m)$ and $\Sigma^{(\beta)}_{bb} (E)=\tilde\Sigma_{(\beta)}(E,\Delta n)$, respectively. And the mutual energy becomes  
\begin{equation}
\Sigma^{(\beta)}_{ab} (E) ={(-\beta)^{\Delta m} e^{-\frac{\Delta m}{\lambda(E)}}}\tilde\Sigma_\beta(E,\Delta n).
\label{MutualEnergyN}
\end{equation}
The energy of BS satisfies the equation
\begin{equation}
E-\delta=\frac{1}{2} \left[ \Sigma^{(\beta)}_{aa}(E)+ \Sigma^{(\beta)}_{bb} (E)\right]
+\frac{\zeta}{2}\sqrt{\Big[\Sigma^{(\beta)}_{aa} (E)- \Sigma^{(\beta)}_{bb} (E)\Big]^2 + 4 \Big[\Sigma^{(\beta)}_{ab} (E)\Big]^2}. 
\label{energyequationN}
\end{equation}
This equation also have up to four real solutions, labeled as $E_{\beta\zeta}$, corresponding to the energies of the BSs. 
The corresponding atomic excitation amplitudes should be described by Eq.~\eqref{uaub}. 
Note that for both states with $E_{\beta,\pm1}$, the photonic excitation amplitudes $f^{(\beta,\pm 1)}_j$ are always \textit{even} functions of $j$, which is different from the separate and braided configurations, where a pair of upper (or lower) BSs have opposite parities. But we can discriminate them by using the sign of $u^{(\beta\zeta)}_a/u^{(\beta\zeta)}_b$, labeled as $\mathrm{sgn}(\tan\Theta_{\beta\zeta})\equiv\eta$. In what follows, we relabel the quantities like $E_{\beta\zeta}, f^{(\beta\zeta)}_j$ as $E_{\beta\eta}, f^{(\beta\eta)}_j$.
\subsection{\label{ThresholdConditions} Threshold conditions and photonic wave functions for the BSs}
\begin{figure*}[t]
\includegraphics[width=\textwidth]{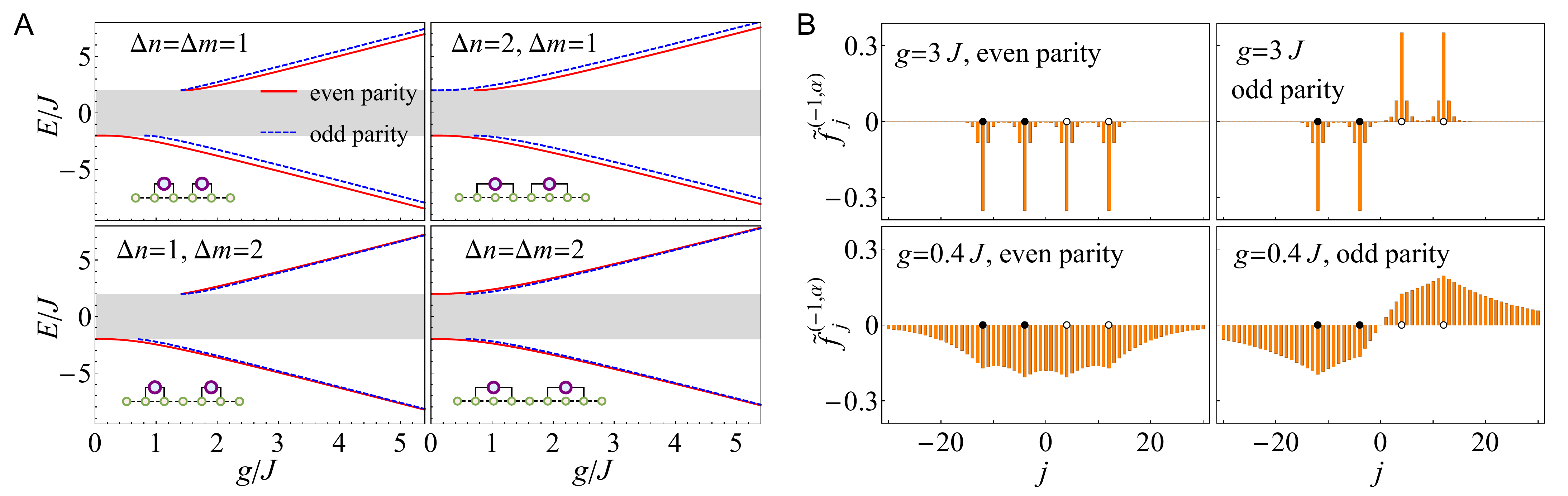}
\caption{(A) The BS energy levels for the case of two separate atoms are plotted as functions of $g$. The inset in each panel shows the corresponding schematic of system. The photonic band is shown by the shaded region.
For all plots $\delta=0$ is assumed. (B) The spatial wave function distribution of photons for two separate giant atoms with distance parameters $\Delta n=\Delta m=8$. The coupling strength is set to $g=3J$ (upper row) and $g=0.4J$ (lower row). The black (white) disks in each panel are used to label the positions of the sites connecting to the atom $a$ ($b$).}
\label{Spectrum2S}
\end{figure*}
\begin{figure*}[t]
\includegraphics[width=\textwidth]{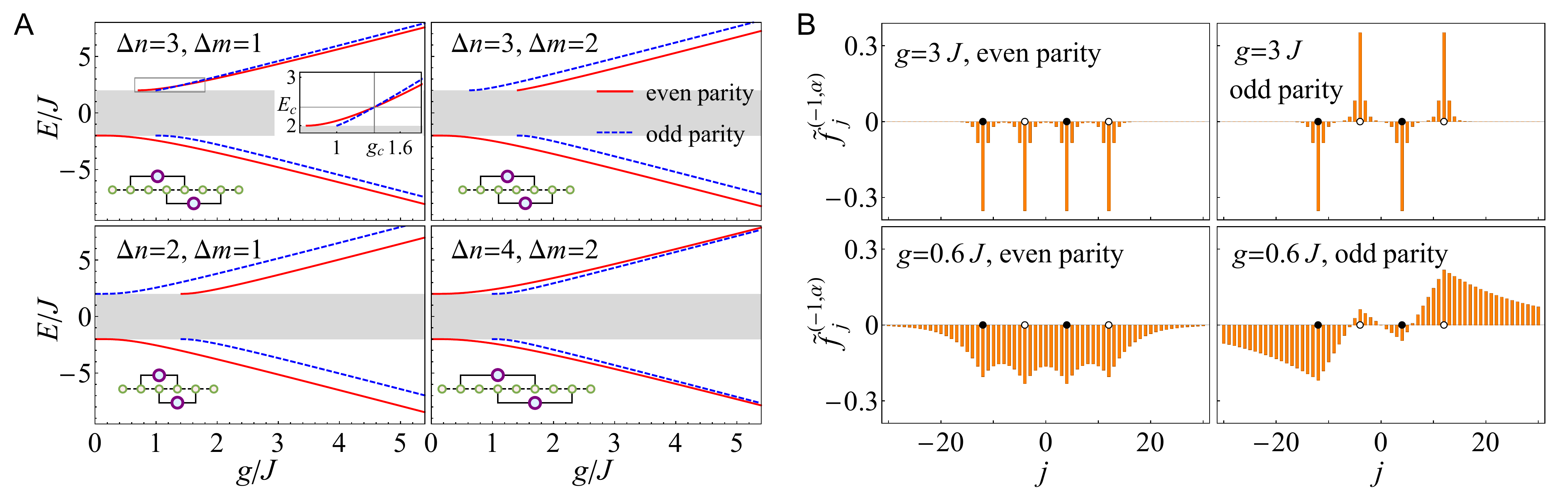}
\caption{(A) The BS energy levels for the case of two braided atoms are plotted as functions of $g$. The inset in each panel shows the corresponding schematic of system. The photonic band is shown by the shaded region.
For all plots $\delta=0$ is assumed. (B) The spatial wave function distribution of photons for two braided giant atoms with distance parameters $\Delta n=16, \Delta m=8$. The coupling strength at each connection points is set to $g=3J$ (upper row) and $g=0.6J$ (lower row). The black (white) disks in each panel are used to label the positions of the sites connecting to the atom $a$ ($b$).}
\label{Spectrum2B}
\end{figure*}
\begin{figure*}[t]
\includegraphics[width=\textwidth]{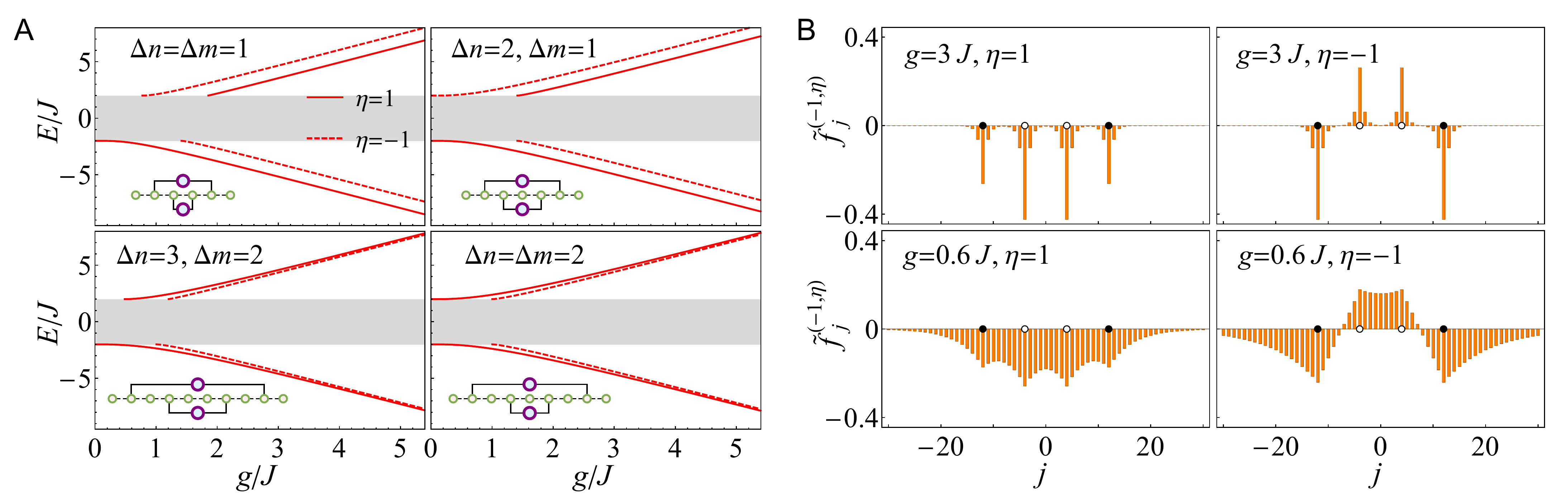}
\caption{(A) The BS energy levels for the case of two nested atoms are plotted as functions of $g$. The inset in each panel shows the corresponding schematic of system. The photonic band is shown by the shaded region.
For all plots $\delta=0$ is assumed. (B) The spatial wave function distribution of photons for two nested giant atoms with distance parameters $\Delta n=\Delta m=8$. The coupling strength at each connection points is set to $g=3J$ (upper row) and $g=0.6J$ (lower row). The left (right) column corresponds to the BS with $\eta=1$ ($\eta=-1$). The black (white) disks in each panel are used to label the positions of the sites connecting to the atom $a$ ($b$).}
\label{Spectrum2N}
\end{figure*}
Like the case of single giant atom, under some parameters, the appearance of the double-atom BS requires a coupling strength larger than a threshold value. By analyzing Eqs.~\eqref{energyequationP} and \eqref{MutualEnergyN}, we can derive the threshold conditions (with $\delta=0$) for the BSs, which are summarized below. 

\textit{Threshold conditions for separate configuration}:

(i) The lower BS with even parity always exists, while another odd-parity one with higher energy exists only when $g>\sqrt{{2J^2}/{(\Delta n+2\Delta m)}}$, as shown in Fig.~\ref{Spectrum2S}A.

(ii) The appearance of upper BSs depends on the parameters $\Delta n$ and $\Delta m$, which can be summarized as follows:
\begin{itemize}
\item
If $\Delta n\in\mathbb{O}^{+}$ and $\Delta m\in\mathbb{O}^{+}$ ($\Delta m\in\mathbb{E}^{+}$), two BSs with opposite parities appear simultaneously when $g>\sqrt{{2J^2}/{\Delta n}}$, and the energy of the odd-parity (even-parity) state is higher than that of the even-parity (odd-parity) one, as shown in the left column of Fig.~\ref{Spectrum2S}A. 
\item
If $\Delta n\in\mathbb{E}^{+}$, and $\Delta m\in\mathbb{O}^{+}$ ($\Delta m\in\mathbb{E}^{+}$), there always exists an upper BS with odd (even) parity, while another upper BS with opposite parity and lower energy only exists when $g>\sqrt{{2J^2}/{(\Delta n+2\Delta m)}}$, as shown in the right column of Fig.~\ref{Spectrum2S}A. 
\end{itemize}

\textit{Threshold conditions for the braided configuration}:

(i) A lower BS with even parity always exists, and another odd-parity one with higher energy appears  when $g>\sqrt{{2J^2}/{(\Delta n-\Delta m)}}$, as shown in Fig.~\ref{Spectrum2B}A.

(ii) The appearance of upper BSs also depends on the parameters $\Delta n$ and $\Delta m$, which can be summarized as follows:

\begin{itemize}
\item If $\Delta n\in\mathbb{O}^{+}$, $\Delta m\in\mathbb{O}^{+}$ ($\Delta m\in\mathbb{E}^{+}$) and $\Delta n>2\Delta m$, there exists an upper BS with parity $\alpha=\pm1$ when $g>\sqrt{{2J^2}/{(\Delta n\pm\Delta m)}}$ [$g>\sqrt{{2J^2}/{(\Delta n\mp\Delta m)}}$], as shown in the upper left panel of Fig.~\ref{Spectrum2B}A. In addition, the two upper BSs are degenerate with $E_{1,1}=E_{1,-1}=E_c$ when $g=\frac{1}{2}e^{\frac{x}{2}}\sqrt{{E_c\sqrt{E_c^2-4J^2}}{\mathrm{csch}(x)}}\equiv g_c$. Here $x={\Delta n}/{[2\lambda(E_c)]}$, $E_c$ is the solution of the transcendental equation $\Sigma^{(1)}_{ab}(E)=0$ in the domain $E>2J$. The energy of the odd-parity (even-parity) state is lower than that of the even-parity (odd-parity) one when $g<g_c$, whereas the result is opposite when $g>g_c$. For the parameters in the upper left panel of Fig.~\ref{Spectrum2B}A, we have $g_c\simeq1.356J$ and $E_c\simeq2.383J$, respectively. 

\item If $\Delta n\in\mathbb{O}^{+}$, $\Delta m\in\mathbb{O}^{+}$ ($\Delta m\in\mathbb{E}^{+}$) and $\Delta n<2\Delta m$, the threshold conditions for the upper BSs are the same as the case of $\Delta n>2\Delta m$. While the energy of the odd-parity (even-parity) state is always lower than the even-parity (odd-parity) one, as shown in the upper right panel of Fig.~\ref{Spectrum2B}A.

\item If $\Delta n\in\mathbb{E}^{+}$, and $\Delta m\in\mathbb{O}^{+}$ ($\Delta m\in\mathbb{E}^{+}$), an upper BS with odd (even) parity always exists, another even-parity (odd-parity) one, with lower energy, only exists when $g>\sqrt{{2J^2}/{(\Delta n-\Delta m)}}$, as shown in the lower row of Fig.~\ref{Spectrum2B}A. 
\end{itemize}

\textit{Threshold conditions for the nested configuration}:

(i) There always exists a lower BS with $\eta=1$, and another one with $\eta=-1$ and higher energy appears when $g>\sqrt{{2J^2}/{\Delta m}}$, as shown in Fig.~\ref{Spectrum2N}A.

(ii) The appearance of upper BSs depends on the parameters $\Delta n$ and $\Delta m$, which can be summarized as follows:
\begin{itemize}
\item If $\Delta n\in\mathbb{O}^{+}$ and $\Delta m\in\mathbb{O}^{+}$ ($\Delta m\in\mathbb{E}^{+}$), an upper BS with $\eta=\mp1$ ($\eta=\pm1$) appears when 
$g>[{2J^2}/{(\Delta n+\Delta m\pm\sqrt{\Delta n ^2+ \Delta m ^2}})]^{\frac{1}{2}},$
and the energy of the state with $\eta=-1$ ($\eta=1$) is higher, as shown in the left column of Fig.~\ref{Spectrum2N}A. 
\item If $\Delta n\in\mathbb{E}^{+}$ and $\Delta m\in\mathbb{O}^{+}$ ($\Delta m\in\mathbb{E}^{+}$), an upper BS with $\eta=-1$  ($\eta=1$) always exists, and another one with lower energy and $\eta=1$  ($\eta=-1$) appears when $g>\sqrt{{2J^2}/{\Delta m}}$, as shown in the right column of Fig.~\ref{Spectrum2N}A. 
\end{itemize}

Figures \ref{Spectrum2S}B, \ref{Spectrum2B}B and \ref{Spectrum2N}B show the photonic wave functions of the lower BSs for three configurations. 
For the separate or braided configurations, the two BSs have opposite parities (see Figs.~\ref{Spectrum2S}B and \ref{Spectrum2B}B). And for the nested configuration, both the two BSs are even parity (see Fig.~\ref{Spectrum2N}B). 
For relatively large coupling strength, the photons are highly localized around the coupling points  (see the upper rows in Figs.~\ref{Spectrum2S}B-\ref{Spectrum2N}B). On the contrary, 
for small coupling strength, the overlap between the single-atom wave-packets becomes remarkable and the mutual distortion of the wave packets 
should be taken into account. As a result, in the relatively weak coupling regime, the even-parity BS for the separate or braided configuration (or the BS with $\eta=1$ for the nested configuration) forms a “bonding” state with the photons being more localized in the area between the two atoms. In contrast, 
the odd-parity BS for the separate or braided configuration (or the BS with $\eta=-1$ for the nested configuration) can be regarded as an “antibonding” state with the photons being more delocalized (see the lower rows in Figs.~\ref{Spectrum2S}B-\ref{Spectrum2N}B). 

\subsection{\label{DDcoupling} Dipole-dipole-like coupling between dressed atoms}
For double giant atoms, the photonic clouds shared by them can produce highly tunable interaction between atoms. Moreover, this result can be generalized to the case of multiple giant atoms, leading to potential applications in quantum simulation and quantum information processing.  
In the strong coupling regime $g\gg J$, the photons are highly localized around the coupling points. Thus only a small overlap exists between the photonic wave functions associated with the single-atom BSs, which can induce a small splitting of the energies centered at $E_{\beta}\simeq \beta\sqrt{2}g$ (assuming $\delta=0$) with $|E_{\beta,1}-E_{\beta,-1}|\ll |E_{\beta}|$ (Here $1$ and $-1$ are used to label the even- and odd-parity BSs, $E_{\beta}$ is the BS energy for a single giant atom.), charactering the dipole-dipole-like coupling between distant dressed states. In this regime, the Hamiltonian can be approximated as 
\begin{equation}
H\simeq E_\beta\sum_{i=a,b} D^\dagger_\beta (i) D_\beta (i) 
+\frac{1}{2}U_{\beta}\left[D^\dagger _\beta (a) D_\beta (b)+\mathrm{H}.\mathrm{c}.\right], 
\label{Heff2Atom}
\end{equation}
where $D_{\beta}(i)$ are the single-atom dressed-state operators, which are treated as mutually commuting degrees of freedom. $U_{\beta}$ is the strength of dipole-dipole-like coupling strength.
This type of interaction can be understood as the emitters exchanging virtual photons through the bath, which are localized around the coupling points. 

(a) For separate configuration [Fig.~\ref{TwoGsys}(a)], the coupling strength is
\begin{equation}
U_{\beta}\simeq\frac{1}{2}(-1)^{\Delta m}\beta^{\Delta m+1}(\sqrt{2}\tilde{g})^{1-\Delta m}J,
\label{UddS}
\end{equation}
where $\tilde{g}={g}/{J}$. In present strong coupling regime,  $U_{\beta}$ is only dependent on $\Delta m$ and not influenced by $\Delta n$, because it is mainly contributed from the overlap of the photonic clouds between the coupling points $n_{a2}$ and $n_{b1}$ (with a distance of $\Delta m$). Furthermore, $U_{\beta}$ should be proportional to the effective coupling strength $\sqrt{2}g$ between a single atom and the waveguide, and meanwhile decrease exponentially with $\Delta m$ (charactered by a factor $e^{-\Delta m/\lambda_{\beta}}$, which can be further approximated as $[J/(\sqrt{2}g)]^{\Delta m}$ in the strong coupling limit). Consequently, for $\Delta m=1$, the trade off between these two effects results in a constant dipole-dipole-like coupling $|U_{\beta}|=J/2$ independent of $g$, thus the splitting between energy levels $E_{\beta,1}$ and $E_{\beta,-1}$ is always $J/2$ for large $g$, as shown in the upper row of Fig.~\ref{Spectrum2S}A. For $\Delta m>1$, this leads to a ($\Delta m-1$)-th power inverse law of $|U_{\beta}|\propto 1/g^{\Delta m-1}$, thus the splitting vanishes rapidly as $g$ increases, resulting in  $E_{\beta,1}\simeq E_{\beta,-1}\simeq E_{\beta}\simeq \beta\sqrt{2}g$ for large $g$, as shown in the lower row of Fig.~\ref{Spectrum2S}A.  

(b) For braided configuration [Fig.~\ref{TwoGsys}(b)], the coupling strength is
\begin{equation}
U_{\beta}\simeq
\left\{
\begin{array}{ll}
\frac{1}{2}(-1)^{\Delta m}\beta^{\Delta m+1}\big(\sqrt{2}\tilde{g}\big)^{1-\Delta m}J,&\Delta n>2\Delta m \vspace{2ex},
\\
(-1)^{\Delta m}\beta^{\Delta m+1}\big(\sqrt{2}\tilde{g}\big)^{1-\Delta n+\Delta m}J,&\Delta n<2\Delta m \vspace{2ex},
\\
\frac{3}{2}(-1)^{\Delta m}\beta^{\Delta m+1}\big(\sqrt{2}\tilde{g}\big)^{1-\Delta m}J,&\Delta n=2\Delta m.
\end{array}
\right.
\label{UddB}
\end{equation}
The factors $1/2$, $1$, and $3/2$ in the above expressions are related to the overlapping degree of the photonic clouds belonging to different atoms. Specifically, when $\Delta n>2\Delta m$, $U_{\beta}$ is mainly contributed from the photonic clouds in the regions $n_{b1}<j<n_{a2}$ with a distance of $\Delta m$, thus the expression of $U_{\beta}$ is the same as that of two separate atoms [see Eqs.~\eqref{UddS} and the first line of Eq.~\eqref{UddB}]. When $\Delta n<2\Delta m$, $U_{\beta}$ is mainly contributed from the photons in two regions $n_{a1}<j<n_{b1}$ and $n_{a2}<j<n_{b2}$, both with width $\Delta n-\Delta m$, resulting in the expression described by the second line of Eq.~\eqref{UddB}. When $\Delta n=2\Delta m$, the photonic clouds in all the three interatomic regions (all with width $\Delta m$) contribute equally, thus we can obtain the third line of Eq.~\eqref{UddB}. Above results can be directly seen from the splitting between energy levels $E_{\beta,1}$ and $E_{\beta,-1}$. Specifically,
(i) for $\Delta m=1, \Delta n>2$, the splitting is approximately $J/2$, as shown in the upper left panel of Fig.~\ref{Spectrum2B}A; (ii) for $\Delta m>1, \Delta n-\Delta m=1$, the splitting is approximately $J$, as shown in the upper right panel of Fig.~\ref{Spectrum2B}A; (iii) for $\Delta m=1, \Delta n=2$, the splitting  is approximately $3J/2$, as shown in the lower left panel of Fig.~\ref{Spectrum2B}A; (iv) for $\Delta m>1, \Delta n-\Delta m>1$, and $\Delta n\geqslant 2\Delta m$ ($\Delta n<2\Delta m$), $|U_{\beta}|\propto 1/g^{\Delta m-1}$ ($|U_{\beta}|\propto 1/g^{\Delta n-\Delta m-1}$) exhibits an inverse power law. Thus the splitting vanishes rapidly as $g$ increases, resulting in $E_{\beta,1}\simeq E_{\beta,-1}\simeq E_{\beta}\simeq \beta\sqrt{2}g$ when $g$ is largre enough, as shown in the lower right panel of Fig.~\ref{Spectrum2B}A.

(c) For nested configuration [Fig.~\ref{TwoGsys}(c)], we have
\begin{equation}
U_{\beta}\simeq(-1)^{\Delta m}\beta^{\Delta m+1}\big(\sqrt{2}\tilde{g}\big)^{1-\Delta m} J,
\label{UddN}
\end{equation}
which is mainly contributed from the photonic clouds in the regions $n_{a1}<j<n_{b1}$ and $n_{b2}<j<n_{a2}$, both with width $\Delta m$. Thus it is not a surprise that the expression of $U_{\beta}$ for this case can be obtained from that for the case of two separate atoms by replacing the factor $1/2$ with $1$  [see Eqs.~\eqref{UddS} and \eqref{UddN}]. 
In addition, for $\Delta m=1$, we have $|U_{\beta}|\simeq J$, indicating that the splitting between energy levels $E_{\beta,1}$ and $E_{\beta,-1}$ is always $J$ for large $g$, as shown in the upper row of Fig.~\ref{Spectrum2N}A. And for $\Delta m>1$, the splitting exhibits a ($\Delta m-1$)-th power inverse law of $|U_{\beta}|\propto 1/g^{\Delta m-1}$, resulting in  $E_{\beta,1}\simeq E_{\beta,-1}\simeq E_{\beta}\simeq \beta\sqrt{2}g$ for large $g$, as shown in the lower row of Fig.~\ref{Spectrum2N}A.  
\section{\label{BS-many-GA}Bound states for multiple giant atoms}
\begin{figure*}[t]
\includegraphics[width=\textwidth]{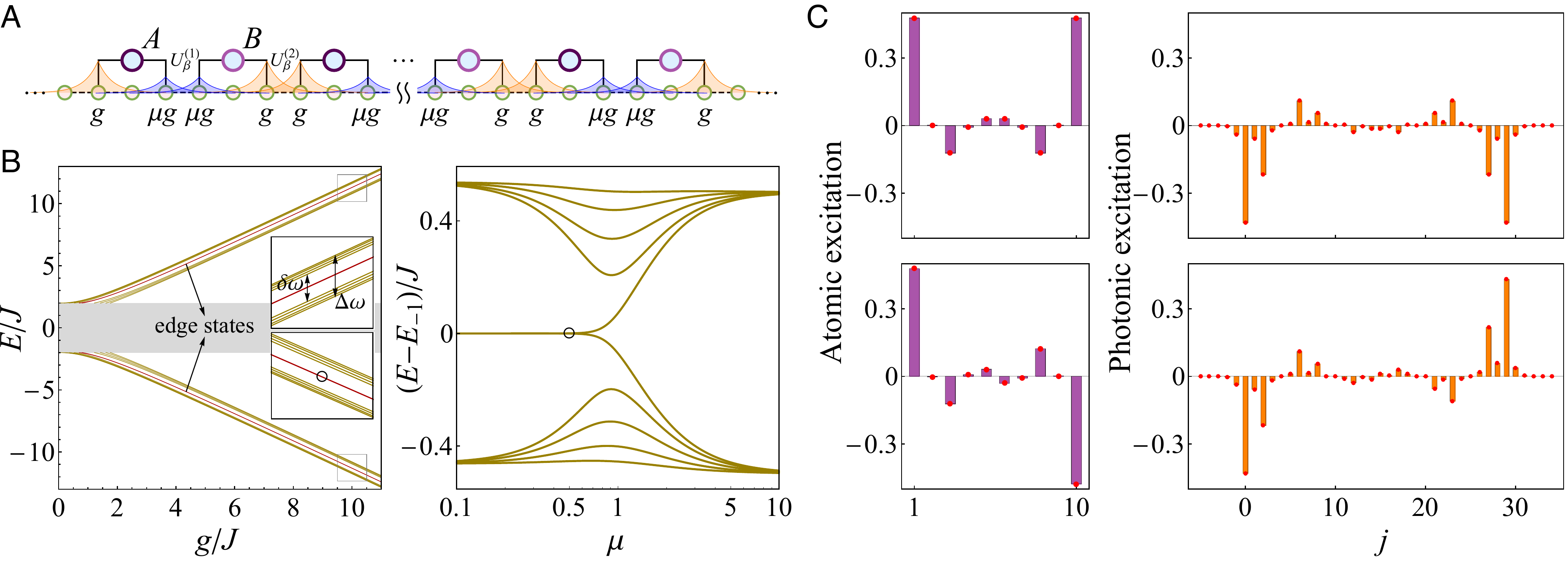}
\caption{(A) Sketch of a setup with separate giant atoms realizing an SSH chain. The distance parameters are set as $\Delta n=2$ and $\Delta m=1$, respectively.
(B) Left panel: the BS energy levels are plotted as functions of the coupling strength $g$ for the setup shown in (A). The parameters are $N_{\mathrm{a}}=10$, $\mu=0.5$, and $\delta=0$. The photonic band is shown by the shaded region. Right panel: the BS energy levels are plotted as functions of $\mu$ for the setup shown in (A). The coupling strength is set as $g=10J$. Other parameters are the same as the left panel.
(C) Left column:  the upper (lower) panel shows the atomic excitation amplitudes for the edge state with even (odd) parity in the lower metaband gap, as indicated by the circles in (B), i.e., $g=10J$ and $\mu=0.5$. 
Right column: the corresponding photonic excitation amplitudes. The bars are numerical results and the red disks are approximate results from the effective Hamiltonian \eqref{H1DTop}.}
\label{TopArraySe}
\end{figure*}

The above analysis can be extended to multiple atoms.  For large $N_{\mathrm{a}}$ and strong atom-waveguide coupling $g$, due to the photon-mediated interactions, the lower (upper) BSs form a metaband structure for propagating dressed-state excitations below (above) the bare photonic band. 
When the coupling strength becomes smaller than a threshold value, a fraction of the dressed-state band merges with the continuum (see Fig.~\ref{SpectrumManyS}A and Fig.~\ref{TopArraySe}B).
Due to the diversity brought about by the distribution of connection points, the metaband structures and their corresponding threshold conditions for multi-giant-atom systems become more complex and tunable than those for small atoms \cite{Calajo-PRA2016}. 
This could make the system a useful platform for 
quantum simulation. In this section, we focus on the 1D array of interacting dressed atoms described by: (i) the normal 1D 
tight-binding model, (ii) the SSH model \cite{Su-PRL1979}. 
\subsection{\label{MB-1DTB}Metaband structures for dressed-atom array described by normal 1D tight-binding model}
\begin{figure*}[t]
\includegraphics[width=\textwidth]{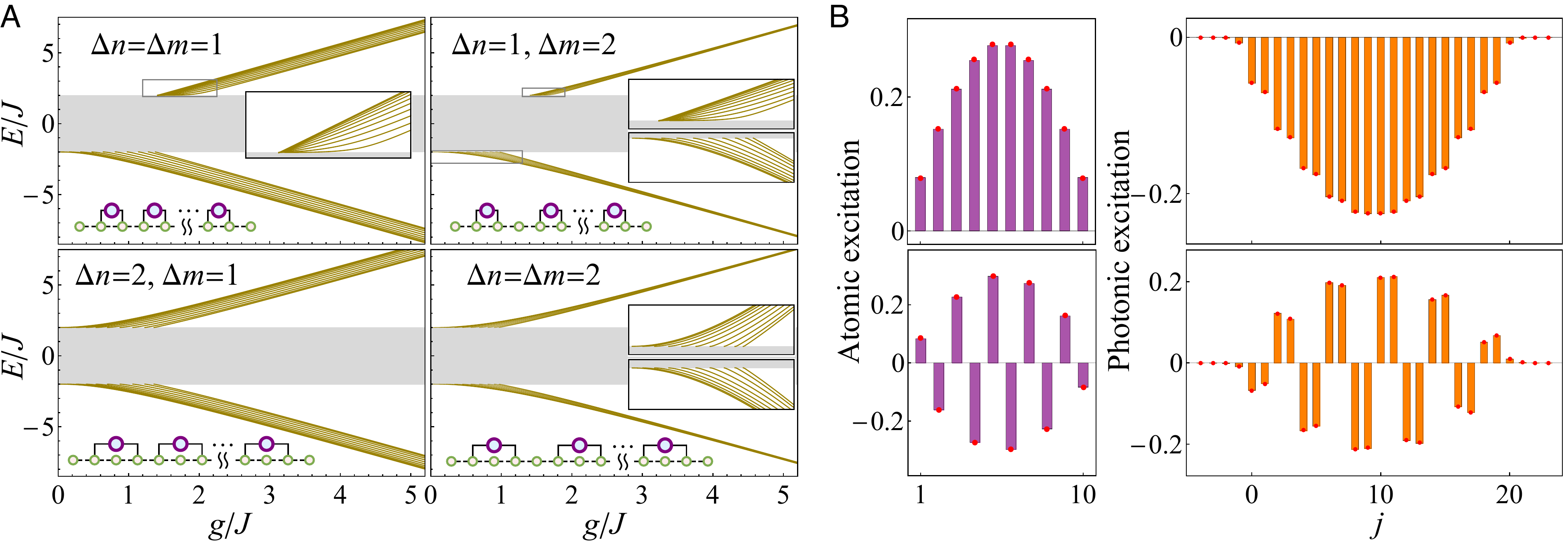}
\caption{(A) The numerically calculated BS energy levels for the case of $N_{\mathrm{a}}=10$ separate atoms are plotted as functions of $g$. The photonic band is shown by the shaded region. The inset in each panel shows the corresponding schematic of system. 
For all plots $\delta=0$ is assumed. (B) Left column:  the atomic excitation amplitudes for the lowermost (uppermost) BS in the lower metaband, as shown in the upper (lower) panel. Right column: the corresponding photonic excitation amplitudes. The bars are numerical results and the red disks are approximate ones from the effective Hamiltonian \eqref{H1DTB}. The parameters are $N_{\mathrm{a}}=10$, $\Delta n=\Delta m=1$, $\delta=0$, and $g=5J$.}
\label{SpectrumManyS}
\end{figure*}
Now we consider a 1D chain of $N_{\mathrm{a}}$ 
atoms, each pair of neighboring atoms is in a separate configuration, with distance parameters $\Delta n$ and $\Delta m$ defined in Sec.~\ref{BS-two-GA}. 
Here we focus on the regime $g\gg J$, where the photon-mediated interactions between atoms are dipole-dipole-like ones. Moreover, according to Eq,~\eqref{UddS}, the ratio of nearest-neighbor coupling to next-nearest-neighbor one can be approximated as $(\sqrt{2}\tilde{g})^{\Delta n+\Delta m}\gg 1$. Thus the next-nearest-neighbor coupling can be neglected, and the array of dressed atoms forms a 1D tight-binding chain with nearest-neighbor hopping rate $U_{\beta}/2$ described by Eq.~\eqref{UddS} (assuming $\delta=0$), the corresponding effective Hamiltonian can be approximated as 
\begin{equation}
H=E_\beta\sum_{i=1}^{N_{\mathrm{a}}} D^\dagger_\beta (i) D_\beta (i)+\frac{1}{2}U_{\beta}\sum_{i=1}^{N_{\mathrm{a}}-1} \left[D^\dagger_\beta (i)D_\beta (i+1) +\text{H.c.} \right].
\label{H1DTB}
\end{equation} 
According to well known results of tight-binding model, the energy spectrum of the Hamiltonian \eqref{H1DTB} exhibits a metaband structure
around the frequency $E_{\beta}\simeq\beta\sqrt{2}g$ [the frequency of a single-atom dressed-state in the strong coupling regime, given in Sec.~\ref{BS-one-GA}] with total width $\Delta\omega\equiv 2|U_{\beta}|\simeq (\sqrt{2}\tilde{g})^{1-\Delta m} J$. 
For $\Delta m=1$, we have $\Delta\omega=J$, i.e., the spectrum width of the metaband for large $g$ is one quarter of the width of photonic band, as shown in the left column of Fig.~\ref{SpectrumManyS}A. And for $\Delta m>1$, the spectrum width as $g$ increases, exhibiting ($\Delta m-1$)-th power inverse law of $\Delta\omega\propto 1/g^{\Delta m-1}$, as shown in the right column of Fig.~\ref{SpectrumManyS}A.  

Figure \ref{SpectrumManyS}B shows the atomic and the photonic excitation amplitudes for the lowermost and the uppermost BSs in the lower metaband, with $\Delta n=\Delta m=1$ and $N_a=10$. The coupling strength is set as a relatively large value $g=5J$, so that the tight-binding Hamiltonian \eqref{H1DTB} is a good approximation. One can see that the results obtained from tight-binding approximation (the red disks in Fig.~\ref{SpectrumManyS}B) are in good agreement with the full numerical ones (the bars in Fig.~\ref{SpectrumManyS}B). The photonic cloud mainly concentrate around the sites connecting to the excited atoms, with the photonic excitation amplitudes being proportional to the corresponding atomic excitation amplitudes. 

\subsection{Topological metaband structures for bound states}
In this subsection, we further provide a proposal to construct a topological spin chain described by the SSH model \cite{Su-PRL1979} based on manipulating dressed-state excitations. This 1D toy model can be used to understand some of the fundamental ideas of topological physics \cite{Hasan-RMP2010,Qi-RMP2011,Haldane-PRL2008,Ozawa-RMP2019}, particularly in low-dimensional systems \cite{Meier-NatureComm2016,Lohse-NaturePhy2016,Nakajima-NaturePhy2016}.

As schematically shown in Fig.~\ref{TopArraySe}A, a 1D chain of $N_\mathrm{a}$ atoms is coupled to the resonator array, with each pair of neighboring atoms in a separate configuration (see Fig.~\ref{TopArraySe}A). The two atoms in a unit cell are labelled as A and B. For the A (B) atom, the coupling strengths at the left and right connection points are $g$ ($\mu g$) and $\mu g$ ($g$), respectively. $\mu$ is a dimensionless positive coefficient. Again, in the regime $g\gg J$, the interactions induced by BS photons
between different atoms are dipole-dipole-like coupling. We also assume that ${\tilde g}^{-\Delta n}\ll\mu$ and $\delta=0$ are satisfied. Thus the strengths of nearest-neighbor coupling are $|U^{(1)}_{\beta}|\simeq{\mu^2 \tilde{g}^{1-{\Delta m}}J}/{(1+\mu^2)^{(\Delta m+1)/2}}$ (for hopping within the unit cell) and $|U^{(2)}_{\beta}|\simeq{\tilde{g}^{1-{\Delta m}}J}/{(1+\mu^2)^{(\Delta m+1)/2}}$ (for hopping connecting neighboring unit cells), respectively.
Moreover, the ratio of nearest-neighbor coupling to next-nearest-neighbor can be much less than $1$, so that the tight-binding approximation is valid. 
As a result, we obtained a topological dressed-atom array described by the SSH model \cite{Su-PRL1979}
\begin{eqnarray}
H=E_\beta\sum_{i=1}^{N_{\mathrm{a}}} D^\dagger_\beta(i)D_\beta(i)+\left[\frac{1}{2}U^{(1)}_{\beta}\sum_{i\in\mathbb{O}^{+}}D^\dagger_\beta(i)D_\beta(i+1)
+\frac{1}{2}U^{(2)}_{\beta}\sum_{i\in\mathbb{E}^{+}}D^\dagger_\beta(i)D_\beta (i+1)+\mathrm{H.c.}\right].
\label{H1DTop}
\end{eqnarray} 
By assuming periodic boundary conditions, one can obtain the corresponding metabands formed by the atom-photon dressed states 
\begin{equation}
E_{\beta,\pm}(K)=E_{\beta}\pm\frac{1}{2}\sqrt{{U^{(1)2}_{\beta}}+{U^{(2)2}_{\beta}}+2U^{(1)}_{\beta}U^{(2)}_{\beta}\cos K},
\end{equation}
where $K\in [-\pi,\pi]$ is the wave vector. From this dispersion relation, one can find that when $\mu\neq 1$ (i.e., $|U^{(1)}_{\beta}|\neq|U^{(2)}_{\beta}|$, note that $|U^{(1)}_{\beta}/U^{(2)}_{\beta}|\simeq\mu^2$ in the strong coupling regime), the spectrum is gapped and forms two symmetric bands centered about the reference frequency $E_{\beta}\simeq\beta\sqrt{1+\mu^2}g$ (the BS energy for a single giant atom), with the spectrum width and band gap 
\begin{subequations}
\begin{equation}
\Delta\omega=\left|U^{(1)}_{\beta}\right|+\left|U^{(2)}_{\beta}\right|\simeq{(1+\mu^2)^{\frac{1-\Delta m}{2}}}{\tilde{g}^{1-{\Delta m}} J},
\end{equation}
\begin{equation}  
\delta\omega=\left|\left|U^{(1)}_{\beta}\right|-\left|U^{(2)}_{\beta}\right|\right|\simeq{(1+\mu^2)^{-\frac{\Delta m+1}{2}}}{\left|1-\mu^2\right|\tilde{g}^{1-{\Delta m}} J}, 
\end{equation}
\end{subequations}
respectively. In particular, for a setup with $\Delta m=1$ (as shown in Fig.~\ref{TopArraySe}A), we have $|U^{(1)}_{\beta}|\simeq \mu^2 J/{(1+\mu^2)}$, $|U^{(2)}_{\beta}|\simeq J/{(1+\mu^2)}$. The 
spectrum width becomes $\Delta\omega\simeq J$, which is independent of $g$. And the band gap becomes $\delta\omega\simeq|1-\mu^2|J/{(1+\mu^2)}$ (see the lower and upper metabands in the left panel of Fig.~\ref{TopArraySe}B). 
With finite systems, the value of $\mu$ determines whether the chain ends with weak hoppings, which leads to the appearance of topologically robust edge states with energy $E_{\beta}$.
As shown by the red lines in the left panel of Fig.~\ref{TopArraySe}B, for $\mu=0.5$ (i.e., $|U^{(1)}_{\beta}|=0.2J,|U^{(2)}_{\beta}|=0.8J$) and $N_{\mathrm{a}}=10$, two degenerate edge states with energy $E_{1}$ ($E_{-1}$) appear in the upper (lower) metaband gap. To show the influence of parameter $\mu$ on the topology of system, 
we plot the BS energy levels of the lower metaband as functions of $\mu$ in the right panel of Fig.~\ref{TopArraySe}B.
We can see that as $\mu<1$ (i.e., $|U^{(1)}_{-1}| < |U^{(2)}_{-1}|$), there are two degenerate edge states with energy $E_{-1}$, indicating that the system is in a topologically non-trivial phase. But, if $\mu>1$ (i.e., $|U^{(1)}_{-1}| > |U^{(2)}_{-1}|$), no edge states appear in the spectrum gap. Therefore, the system is in a topologically trivial phase. In the case of $\mu=1$ (i.e., $|U^{(1)}_{-1}|=|U^{(2)}_{-1}|$) the gap closes, recovering the normal 1D tight-binding model discussed in Sec.~\ref{MB-1DTB}. 

Figure \ref{TopArraySe}C shows the atomic and the photonic excitation amplitudes for the topologically robust edge BSs in the metaband gap (indicated by the hollow circles in Fig.~\ref{TopArraySe}B, corresponding to $g=10J$ and $\mu=0.5$). The atom number is even, with $N_{\mathrm{a}}=10$. One can see that the two degenerate edge states are localized at the ends of the atom array, with even and odd parities, respectively. The results obtained from the approximate Hamiltonian \eqref{H1DTop} (red disks) are in good agreement with the full numerical ones (bars) obtained from the Hamiltonian \eqref{SysHamiltonian}.
Note that for the usual SSH-type atomic chain, the neighboring atoms are directly coupled, thus the eigen states of system contain only excitations
of bare atoms. However, for present system, the dressed atoms, instead of bare atoms, are effectively coupled through photon-mediated interactions, forming collective dressed states. Thus the edge states shown in Fig.~\ref{TopArraySe}C include both atomic and photonic excitations. 
The photonic clouds accompanying the most excited atoms are mainly concentrated around the left and right ends of the atom array.
\section{\label{conclusion}conclusion}
In summary, we analyze in detail the BSs in the single-excitation subspace of a 
system of multiple giant atoms coupled to a common structured photonic bath. The most general analytical 
expressions possible for these states and the corresponding energy spectra are obtained. Based on these, we 
obtain some essential properties of these states. These results reflect unconventional light-matter interactions when giant-atom systems are coupled to a structured environment. Specifically, 
the threshold conditions for the appearance of the BSs and the photon-mediated interactions between atoms for different configurations are analyzed. 
For large atom number and strong atom-photon coupling, the BSs in the photonic band gap can form different types of metaband structures (e.g., SSH-type energy band with nontrivial topological properties), depending on the arrangement of the coupling points. These results can serve as a starting point to further explore other nontrivial many-body phases, making the system a useful platform for quantum simulations and quantum information processings.
\begin{acknowledgments}
This work was supported by the National Natural Science Foundation of China (NSFC) under Grants No. 61871333.
\end{acknowledgments}
\bibliography{MS-March-2-2024YMT}
\end{document}